\documentclass{emulateapj}
\usepackage{epsf}
\usepackage{apjfonts}
\usepackage{xspace}
\usepackage{amsmath}
\usepackage{framed}
\usepackage{color}
\def\H2{{{\rm H}_2}}
\def\HI{{\rm H\,I}}
\def\HII{{\rm H\,II}}

\def\Msun{\, {\rm M}_{\odot}}

\def\dim#1{\mbox{\,#1}}

\def\hide#1{}

\begin{document}

\title{On the Baryonic Contents of Low Mass Galaxies}

\author{Nickolay Y.\ Gnedin\altaffilmark{1,2,3}}  
\altaffiltext{1}{Particle Astrophysics Center, 
Fermi National Accelerator Laboratory, Batavia, IL 60510, USA; gnedin@fnal.gov}
\altaffiltext{2}{Department of Astronomy \& Astrophysics, The
  University of Chicago, Chicago, IL 60637 USA} 
\altaffiltext{3}{Kavli Institute for Cosmological Physics and Enrico
  Fermi Institute, The University of Chicago, Chicago, IL 60637 USA}

\begin{abstract}
The baryonic Tully-Fisher relation is an important observational
constraint on cosmological and galactic models. However, it is
critical to keep in mind that in observations only stars, molecular,
and atomic gas are counted, while the contribution of the ionized gas
is almost universally missed. The ionized gas is, however, expected to
be present in the gaseous disks of dwarf galaxies simply because they
are exposed to the cosmic ionizing background and to the stellar
radiation that manages to escape from the central regions of the
galactic disks into their outer layers. Such an expectation is,
indeed, born out both by cosmological numerical simulations and by
simple analytical models.
\end{abstract}

\keywords{galaxies: kinematics and dynamics -- galaxies: dwarf -- galaxies: irregular -- galaxies: spiral -- methods: numerical}

\section{Introduction}
\label{sec:intro}

The baryonic Tully-Fisher relation
\citep[BTFR;][]{gals:f99,gals:msbd00} extends the classical
Tully-Fisher relation \citep{gals:tf77} by including, in addition to
stars, all the baryonic content of galaxies. The physical reason for
such an extension is very sound - in galactic halos that maintain the
universal fraction of baryons, the baryonic mass becomes a linear
proxy of the total mass. A deviation from the expected total mass -
rotational velocity relation may indicate missing baryons \citep[although
such deviations may be degenerate with the effect of stellar
feedback, e.g.][]{gals:dv09}; hence
baryonic Tully-Fisher relation becomes a powerful test of galaxy
formation models.

It is not surprising, therefore, that a significant effort has been
expended over the years in measuring the BTFR in a diverse set of
galaxies \citep{gals:v01,gals:bj01,gals:gmfj04,gals:m05,gals:pr05,
gals:bcks08,gals:sms09,gals:tbmh09,gals:gfjs10,gals:teap11}. The
current state of affairs is well synthesized by \citet{gals:m11b}. The
remarkable power-law behavior of BTFR over 4 decades in mass indicates
a substantial fraction of ``missing'' baryons in the lowest mass
galaxies. The observed ``baryonic'' components of these galaxies are
dominated by atomic gas, and the decreasing ``baryonic'' fraction with
galaxy mass is often interpreted as a substantial mass loss from the
dwarf galaxies.

A limitation of such an argument, however, is that no observation
actually measures the \emph{baryonic} contents of galaxies; only
stars, molecular, and atomic gas are accounted in observational
studies, but not the ionized gas (with a handful of exceptions). In
other words, in dwarf, $\HI$ dominated galaxies, the observed BTFR is,
in fact, an $\HI$ Tully-Fisher relation. One needs, therefore,
be careful to distinguish the total BTFR that accounts for all of the
baryons within the virial radius of a galaxy, and the observed BTFR
that only includes stars and molecular and atomic gas, but does not
account for the ionized gas.

In this paper I demonstrate that dwarf galaxies are expected to
contain warm ionized gas (to be distinguished from hot halos) - simply
because the gaseous disks of galaxies are exposed to the cosmic
background radiation (plus whatever of their own ionizing radiation
escapes into the outer layers of their disks) , which is going to
ionize the outer layer of atomic disks down the column density $N_{\rm
H}\sim10^{19}\dim{cm}^2$, comparable to the transition column density
between Lyman Limit systems (which are mostly ionized) and Damped
Lyman-$\alpha$ systems (which are mostly neutral).

Hence, the interpretation of the observed BTFR is rather non-trivial,
nor can it easily be used to deduce the fraction of baryons ejected by
the feedback from dwarf galaxies.

\section{Simulations}
\label{sec:sims}

The simulation used in this paper is similar to the one
described in \citet{ng:gk10}. Specifically, the Adaptive Refinement
Tree (ART) code \citep{misc:k99,misc:kkh02,sims:rzk08} is employed to
model a $6h^{-1}\dim{Mpc}$ cube centered on a typical $~2L_*$ (at
$z=0$) galaxy. The Lagrangian region of a sphere with radius equal to
$5R_{\rm vir}$ at $z=0$ (a ``region of interest'') is sampled in the
initial conditions with the effective $512^3$ resolution, while the
rest of the simulation volume is sampled more crudely. This setup
results in the mass resolution in the region of interest of
$1.3\times10^6\Msun$ in dark matter. This region of interest is then
allowed to refine adaptively as the simulation proceeds in a
quasi-Lagrangian manner, all the way down to additional 6 levels of
refinement (total spatial dynamic range of about 30{,}000),
maintaining spatial resolution of $260\dim{pc}$ in comoving reference
frame. The simulation adopts $\Lambda$CDM cosmology similar to the
WMAP1 one ($\Omega_M=0.3$, $\Omega_B=0.046$, $\sigma_8=0.9$ and
$h=0.7$).

The physical processes modeled in the simulation are exactly the same
as described in \citet{ng:gk10}, with one exception. In particular,
the simulation incorporates gas cooling (including cooling on metals, molecular
hydrogen, and dust), a phenomenological model for molecular hydrogen
formation, full time-dependent and spatially variable 3D radiative
transfer of ionizing and Lyman-Werner band radiation (both from local
sources and from the incident cosmic background of \citet{jnu:hm01})
using the Optically Thin Variable Eddington Tensor (OTVET)
approximation of \citet{ng:ga01}, and star formation in the molecular
gas using \citet{sfr:kt07} recipe. The only difference from the
\citet{ng:gk10} simulation is that the supernova feedback is disabled.

Existing models of supernova feedback in cosmological simulations (and
other types of stellar feedback, except the feedback from stellar
ionizing radiation, that is explicitly included) are still
insufficiently developed to be fully realistic. Therefore, I
deliberately disable that form of feedback to limit numerical
treatment to only those physical processes that can be modeled with a
reasonable degree of realism.

It is difficult to envision that the supernova feedback would
\emph{increase} the baryonic mass of galaxies, hence baryonic
fractions in model galaxies should be considered as upper limits
on the baryonic fractions in a real universe; any reasonable stellar
feedback models should reduce these fractions even further.

In order to verify that the results presented below are robust and
numerically converged, I re-run the last $150\dim{Myr}$ of the
time evolution in the simulation with an additional level of
refinement (twice higher spatial resolution) and more aggressive
refinement criteria (to enlarge the regions refined to the highest
allowed level). This convergence test has demonstrated that all the
results presented below are highly robust.

\section{Total and Observed Baryonic Tully-Fisher Relations}
\label{sec:tfrs}

\begin{figure}[t]
\plotone{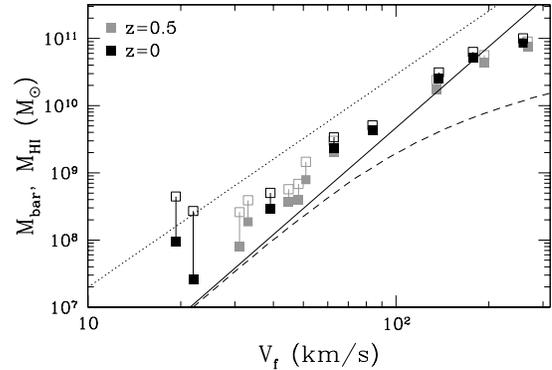}
\caption{The baryonic Tully-Fisher relation for model galaxies at
  $z=0$ (black) and $z=0.5$ (gray); I show two redshifts since the
  $z=0$ sample is very small by construction. Each model galaxy is
  represented by a vertical line with two symbols: open square shows
  the total baryonic mass of the galaxy; filled square shows the
  observed ``baryonic'' mass, i.e.\ the sum of stellar mass, molecular
  gas mass, and atomic gas mass. The difference between the two
  squares is the contribution of the ionized gas. The solid line is the
  best-fit observed BTFR from \citet{gals:m11b}; the dotted line is
  $f_{\rm uni}M_{200}$ vs $V_{\rm max}$ for dark matter halos in the
  $\Lambda$CDM cosmological model, while the dashed line is $M_{\HI}$
  vs $V_{\rm max}$ obtained by abundance matching of the theoretical
  halo mass function and the observed $\HI$ mass function (see text
  for details).}
\label{fig:mbarvf}
\end{figure}

In order to compare the simulation to the observations, the total
baryonic mass (all the way to the virial radius) and the observed
``baryonic'' mass (i.e.\ the sum of the stellar mass, molecular gas
mass, and atomic gas mass) are measured for all model galaxies from
the region of interest. To measure the rotational velocity, the mass
density profiles for various mass components of model galaxies are
constructed, and the circular velocity $V_f$ are measured at the
radius containing 90\% of all $\HI$. This procedure is similar to the
one adopted by \citet{gals:m11b}. I have verified that, if the radius
including either 80\% or 95\% of all atomic gas is used instead, the
results would change by no more than a symbol size in Figure
\ref{fig:mbarvf}.

Since the region of interest is centered on a single large galaxy,
only a handful of other noticeable isolated galaxies end up in that
region by $z=0$. To somewhat enlarge the sample, I show in Figure
\ref{fig:mbarvf} both $z=0$ and $z=0.5$ simulation snapshots, which
are apparently fully consistent with each other.

Figure \ref{fig:mbarvf} shows the two BTFR (the total baryonic
one and the observed BTFR that only includes stellar mass, molecular
gas mass, and atomic gas mass but not the ionized gas mass). The
solid line tracks the best-fit observed BTFR \citep{gals:m11b}
relation in the form
\[
  M_{\rm bar} = \frac{47\Msun}{(\dim{km/s})^4} V_f^4.
\]
The dotted line gives a relation between the product of the
universal baryon fraction $f_{\rm uni}=\Omega_B/\Omega_M$ and
$M_{200}$ (mass within the overdensity 200 with respect to the
critical density) vs $V_{\rm max}$ for dark matter halos from Bolshoi
and Multi-Dark simulations \citep{cosmo:pkcb11}. This relation is
slightly steeper than $f_{\rm uni}M_{200} \propto V_{\rm max}^3$ due
to the halo concentration dependence on halo mass. The simulation lies
to the right of that line in Figure \ref{fig:mbarvf} because $V_{\rm
max}$ from pure N-body simulations is not a sufficiently accurate
proxy for the disk circular velocity - halo contraction due to
baryonic condensation results in $V_f \approx (1.2-1.3)V_{\rm
max}$.

It is also possible to estimate the $\HI$ masses of dark matter halos
using abundance matching - i.e.\ identifying dark matter halos from a
theoretically computed halo mass function with $\HI$ galaxies from the
actual observed $\HI$ mass function \citep{igm:zmsw05} that have the
same spatial number density. The details on how this matching is
performed are described in \citet{ng:mgsv10} or \citet{sims:tkpr11};
$\HI$ mass vs $V_{\rm max}$ that results from such abundance matching
is shown in Fig.\ \ref{fig:mbarvf} with the dashed line. Not
surprisingly, it matches the best-fit observed BTFR at low masses very
well.

The observed BTFR in the simulation (solid squares) is not far from
the best-fit line to the observational data, but is above it by about
a factor of 2 for smallest galaxies. It is important to remember that
the simulation should only be treated as an upper limit to the
baryonic content of galaxies, because the stellar feedback is
deliberately disabled in the simulation (except the feedback of
ionizing radiation, which is included in the simulation by virtue of
following the full spatially-variable and time-dependent 3D radiative
transfer). The fact that the simulation points are not falling on the
observed relation for $V_f \la 200\dim{km/s}$ simply illustrates the
well-known fact that the stellar feedback processes are important for
determining the baryonic contents of sub-$L_*$ galaxies.

\begin{figure}[t]
\plotone{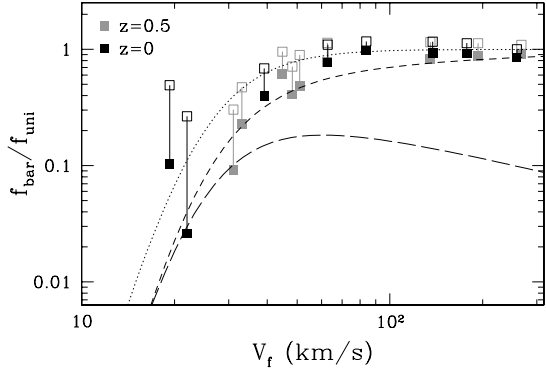}
\caption{The baryonic fraction (in units of the universal fraction
  $f_{\rm uni}=\Omega_B/\Omega_M$) for model galaxies at $z=0$ (black)
  and $z=0.5$ (gray). Symbols are the same as in Figure
  \ref{fig:mbarvf}. The dotted line approximately traces the effect of
  photo-evaporation of gas from halos due to cosmic background
  (Equation \ref{eq:evap}) while the short-dashed line shows a simple
  model (\ref{eq:ion}) for the neutral content of galactic disks in
  dwarf galaxies. The long-dashed line is the fraction of ionized gas
  within the $\HI$ radius (Equation \ref{eq:fion}).}
\label{fig:fbarvf}
\end{figure}

In order to illustrate the crucial difference between the total
baryonic fraction and the observed (stars + molecular + atomic gas)
baryonic fraction in model galaxies, I show the two fractions in
Figure \ref{fig:fbarvf}. As one can see, the total baryonic fraction
in model galaxies remains close to universal all the way to $V_f
\sim 40\dim{km/s}$, and at lower $V_f$ the effect of photo-evaporation
of gas from dark matter halos due to heating by the cosmic ionizing
background radiation becomes important. That effect in this simulation
is approximately described by a fitting formula of \citet{ng:g00b},
\begin{equation}
  \frac{f_{\rm bar}}{f_{\rm uni}} = \left[1 +
  (2^{\alpha/3}-1)\left(\frac{M_c}{M_{200}}\right)^\alpha\right]^{-3/\alpha}
  \label{eq:evap}
\end{equation}
with parameter values $\alpha=1$ and $M_C=7\times10^9\Msun$. The
latter value is the same as found by \citet[][converted from the
virial overdensity of 97 with respect to the critical used there to
the overdensity of 200 used here, for the halo concentration of
14]{dsh:ogt08}.\footnote{I also notice that several recent studies of
the photoionization effect that did not treat the radiative transfer
of ionizing radiation \citep{dsh:hygs06,dsh:cefj07,dsh:ogt08} found
that the best-fit value of $\alpha = 2$. However, in the simulations
with the full spatially-variable and time-dependent 3D radiative
transfer we always find that the best-fit value of $\alpha$ is 1, even
in simulations run with widely different codes, like the moving mesh
code of \citet{ng:g95} and the currently used ART code.} This function
(plotted against $V_{\rm max}$) is shown in Figure \ref{fig:fbarvf} as
a dotted line. In agreement with \citet{gals:mw10}, this line alone is
unable to explain the baryonic contents of dwarf galaxies.

It is also possible to construct a simple analytical model that
approximately reproduces these results for the observed BTFR, following
the ideas of \citet{gals:mmw98}. Let's consider an exponential gaseous
disk with the surface density run
\[
  \Sigma(R) = \Sigma_0 e^{-R/R_d},
\]
such that the disk scale length is a given fraction $\mu$ of the halo
virial radius $R_{200}$,
\[
  R_d = \mu R_{200},
\]
and the disk mass is a fraction $\nu$ of the total halo mass of
$M_{200}$,
\[
  2\pi\Sigma_0 R_d^2 = \nu M_{200},
\]
so that
\[
  \Sigma_0 = \frac{\nu}{\mu^2} \frac{M_{200}}{2\pi R_{200}^2}
\]
(expressions for $\mu$ and $\nu$ are given in \citet{gals:mmw98} as
functions of other parameters). The disk is exposed to the cosmic
ionizing background, which ionizes hydrogen (from both sides) in a
slab of gas below the surface density $\Sigma_\HII$. Then the mass of the
neutral gas in the disk is
\begin{equation}
  M_\HI = 2\pi \int_0^{R_\HII} (\Sigma(R)-\Sigma_\HII) R\, dR,
  \label{eq:mhidisk}
\end{equation}
where $R_\HII$ is the edge of the $\HI$ disk ($\Sigma(R_\HII) =
\Sigma_\HII$). The minus $\Sigma_\HII$ term under the integral in
Equation (\ref{eq:mhidisk}) appears because both sides of the disk are
exposed to the ionizing radiation, and hence each side has an ionized
layer of thickness $\Sigma_\HII/2$. Integral \ref{eq:mhidisk} can be
easily taken, so that the fraction of the disk that is neutral is
\begin{equation}
  f_\HI = \frac{M_\HI}{\nu M_{200}} = 1 -
  \frac{\Sigma_\HII}{\Sigma_0}\left(1+\ln\frac{\Sigma_0}{\Sigma_\HII}+\frac{1}{2}\left(\ln\frac{\Sigma_0}{\Sigma_\HII}\right)^2\right).
  \label{eq:ion}
\end{equation}
This dependence (for the combination $\nu/\mu^2=8$, which is somewhat
on a lower side but still plausible, and $\Sigma_\HII =
0.4\Msun/\dim{pc}^{2}$ - the reason for the latter choice will become
clear below\footnote{This is also the column density of a typical
sub-DLA absorption system.}), multiplied by Equation (\ref{eq:evap})
and plotted as a function of $V_{\rm max}$, is shown in Figure
\ref{fig:fbarvf} with the dashed line. It does not fit the simulation
results perfectly, but roughly traces the lower envelope of filled
symbols; this simplistic model does not account for stellar
contribution to the observed BTFR, and is therefore can only serve as
an approximation to the lower limit of the observed baryonic fraction
of model galaxies.

From an observational perspective, it may be useful to know what
fraction of the ionized gas is actually inside the $\HI$ disk. In the
simple model this quantity is easy to compute,
\[
  M_\HII(R<R_\HII) = 2\pi \int_0^{R_\HII} \Sigma_\HII R\, dR,
\]
or
\begin{equation}
  f_\HII(R<R_\HII) = \frac{M_\HII(R<R_\HII)}{\nu M_{200}} = 
  \frac{1}{2}\frac{\Sigma_\HII}{\Sigma_0}\left(\ln\frac{\Sigma_0}{\Sigma_\HII}\right)^2.
  \label{eq:fion}
\end{equation}
The fraction of ionized gas outside the $\HI$ disk is then simply one
minus the sum of Equations (\ref{eq:ion}) and Equation
(\ref{eq:fion}). The long-dashed line in Figure \ref{fig:fbarvf} shows
Equation (\ref{eq:fion}) as a function of $V_{\rm max}$. The ionized
gas starts dominating the disk mass only for $V_{\rm
  max}<30\dim{km/s}$, but makes a $>20\%$ contribution all the
way to $V_{\rm max}\approx 100\dim{km/s}$.

\begin{figure}[t]
\plotone{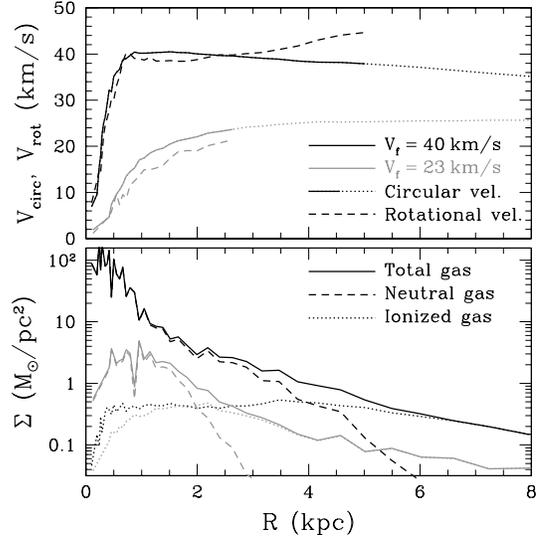}
\caption{The circular and rotational velocity profiles (top) and surface
  density profiles for the total, ionized, and neutral gas (bottom)
  for two model galaxies at $z=0$, one with $V_f=40\dim{km/s}$ (black
  lines) and another one with $V_f=23\dim{km/s}$ (gray lines; the
  lowest of two black symbols at this velocity in Figures
  \ref{fig:mbarvf} and \ref{fig:fbarvf}). In the circular velocity
  plot, the solid segment of the circular velocity shows its extent
  over the $\HI$ disk, whereas the dotted line shows the extension of
  the line into the region where there is no atomic gas.}
\label{fig:profs}
\end{figure}

Finally, one may ask where the ionized gas, apparent in Figures
\ref{fig:mbarvf} and \ref{fig:fbarvf}, is actually located in model
galaxies? To answer that question, I show in Figure \ref{fig:profs}
surface density profiles for the total, ionized, and neutral (atomic
and molecular, although the molecular fraction in the shown galaxies
is small) gas in two galaxies, in which the ionized gas contribution
is large. In both galaxies neutral gas forms an approximately
exponential disk, while the surface density of the ionized component
remains approximately constant well outside the scale length of the
neutral disk. These trends are consistent with the simple analytical
model presented above. The choice of $\Sigma_\HII =
0.4\Msun/\dim{pc}^2$ is the actual value of the fixed surface density
of the ionized gas found in the simulation.

The top panel of Fig.\ \ref{fig:profs} also shows the circular
velocity profiles for the two galaxies and the rotational velocity of
$\HI$ gas. The difference between the two is due to non-circular
motions in the gas; this difference is consistent with the prior
theoretical models \citep{gals:vrkg07,gals:ds10,sims:tkpr11} and
observational measurements of galactic rotation curves
\citep[c.f.][]{gals:tbwb08,gals:agbf11}.

\section{Conclusions}
\label{sec:concl}

The baryonic Tully-Fisher relation is an important observational
constraint on cosmological and galactic models. However, it is
critical to keep in mind that in observations only stars, molecular,
and atomic gas are counted, while the contribution of the ionized gas
is almost universally missed. Hence the observed BTFR does not count
\emph{all baryons}. Comparison of such observations to theoretical
predictions is, therefore, highly non-trivial, and requires a proper
modeling of radiative transfer of ionizing radiation, at least in an
approximate form.

In this paper I present an example of such modeling in the form of a
cosmological numerical simulation with radiative transfer. The
simulation is not fully realistic, since it does not include a
treatment of supernova feedback, and thus overestimates the baryonic
fraction of model galaxies. Even with this incomplete treatment, low
mass model galaxies contain large, and for $V_f \la 50\dim{km/s}$
dominant, contribution of ionized gas. 

The ionized gas is present in the gaseous disks of model galaxies in
two regions: the outer parts of the disks are ionized since their
surface densities (and, hence, column densities) are too low. Even
more importantly, the outer layers of the inner disks are also
ionized, because they are exposed to the cosmic ionizing background
and to the stellar radiation that manages to escape from the central
regions of the galactic disks into their outer layers. These layers
are direct analogs of the Reynolds' layer observed in the Milky Way
galaxy \citep{gals:r93}, but their relative contribution to the total
mass budget becomes progressively larger as galactic disks become less
massive, less dense, and allow ionizing radiation to reach deeper.

The existence of this ionized gas is not just a theoretical conjecture
- it is unavoidable from purely physical grounds, since the cosmic
ionizing background must ionize the outer layers of galactic $\HI$
disks (on both sides) down to column densities of Lyman Limit systems,
$N_\HI \ga 10^{19}\dim{cm}^{-2}$ (in my simulation it is $\Sigma_\HII
= 0.2\Msun/\dim{pc}^2$ or, equivalently, $\HI =
2.5\times10^{19}\dim{cm}^{-2}$).

\acknowledgements 

I am grateful to Andrey Kravtsov for enlightening discussions and
constructive criticism. Throughout the stormy refereeing process,
different referees offered constructive and not-so-constructive
criticisms. I am grateful to all of them, as the final manuscript ended
up being a major improvement over the original draft. This work was
supported in part by the DOE at Fermilab, by the NSF grant
AST-0908063, and by the NASA grant NNX-09AJ54G. The simulations used
in this work have been performed on the Joint Fermilab - KICP
Supercomputing Cluster, supported by grants from Fermilab, Kavli
Institute for Cosmological Physics, and the University of Chicago.
This work made extensive use of the NASA Astrophysics Data System and
{\tt arXiv.org} preprint server.

\bibliographystyle{apj}
\bibliography{ng-bibs/self,ng-bibs/sims,ng-bibs/gals,ng-bibs/misc,ng-bibs/sfr,ng-bibs/cosmo,ng-bibs/dsh,ng-bibs/jnu,ng-bibs/igm}

\end{document}